\documentclass[fleqn,12pt]{wlscirep}
\usepackage[utf8]{inputenc}
\usepackage[T1]{fontenc}
\title{Constraint on the stem cell numbers and division rates posed by the risk of cancer}

\author[1,2*]{Augusto Gonzalez}
\author[3]{Teresita Rodriguez}
\author[1,3]{Rolando Perez}
\affil[1]{University of Electronic Science and Technology, Chengdu, People Republic of China}
\affil[2]{Institute of Cybernetics, Mathematics and Physics, Calle E 309, Havana, CP 10400, Cuba}
\affil[*]{Corresponding author. Email: agonzale@icimaf.cu}
\affil[3]{Center for Molecular Inmunology, Havana, Cuba}%//Affiliation, city, postcode, country


\begin{abstract}
Compiled data for the stem cell numbers, $N_s$, and division rates, $m_s$, is reanalized in order to show 
that we can distinguish two groups of human tissues. In the first one, there is a relatively high fraction 
of maintenance (stem and transit) cells in the tissue, but the division rates are low. The second group, 
on the other hand, is characterized by very high transit cell division rates, of around one division
per day. These groups do not have an embrionary origin. We argue 
that their properties arise from a combination of the needs of tissue homeostasis (in particular turnover 
rate) and a bound on cancer risk, which is roughly a linear function of the product $N_s\times m_s$. The  
bound on cancer risk leads to a threshold at $m_s\approx 8/year$, where the fraction of stem cells falls 
down two orders of magnitude.
\end{abstract}
\begin{document}

\flushbottom
\maketitle
\thispagestyle{empty}

%\noindent Please note: Abbreviations should be introduced at the first mention in the main text – no abbreviations lists. 

\section*{Cancer risk and the product $N_s\times m_s$ in a tissue}

%The Introduction section, of referenced text\cite{Figueredo:2009dg} expands on the background of the work (some overlap with the Abstract is acceptable). The introduction should not include subheadings.

The purpuse of the present paper is to show that the risk of cancer may pose a 
constraint on the number, $N_{s}$, and replication rate, $m_{s}$, of stem cells in a tissue. The
starting point is a noticed correlation \cite{Tomasetti1,Tomasetti2} between the risk and the product 
$N_{s}\times m_{s}$. To the best of our knowledge, the only explicit expression for the risk involving 
this product was obtained in a previous paper of ours \cite{LevyCancer}. For completeness, we briefly 
sketch the main steps leading to that expression.

\begin{figure}[t]
\begin{center}
\includegraphics[width=0.9\linewidth,angle=0]{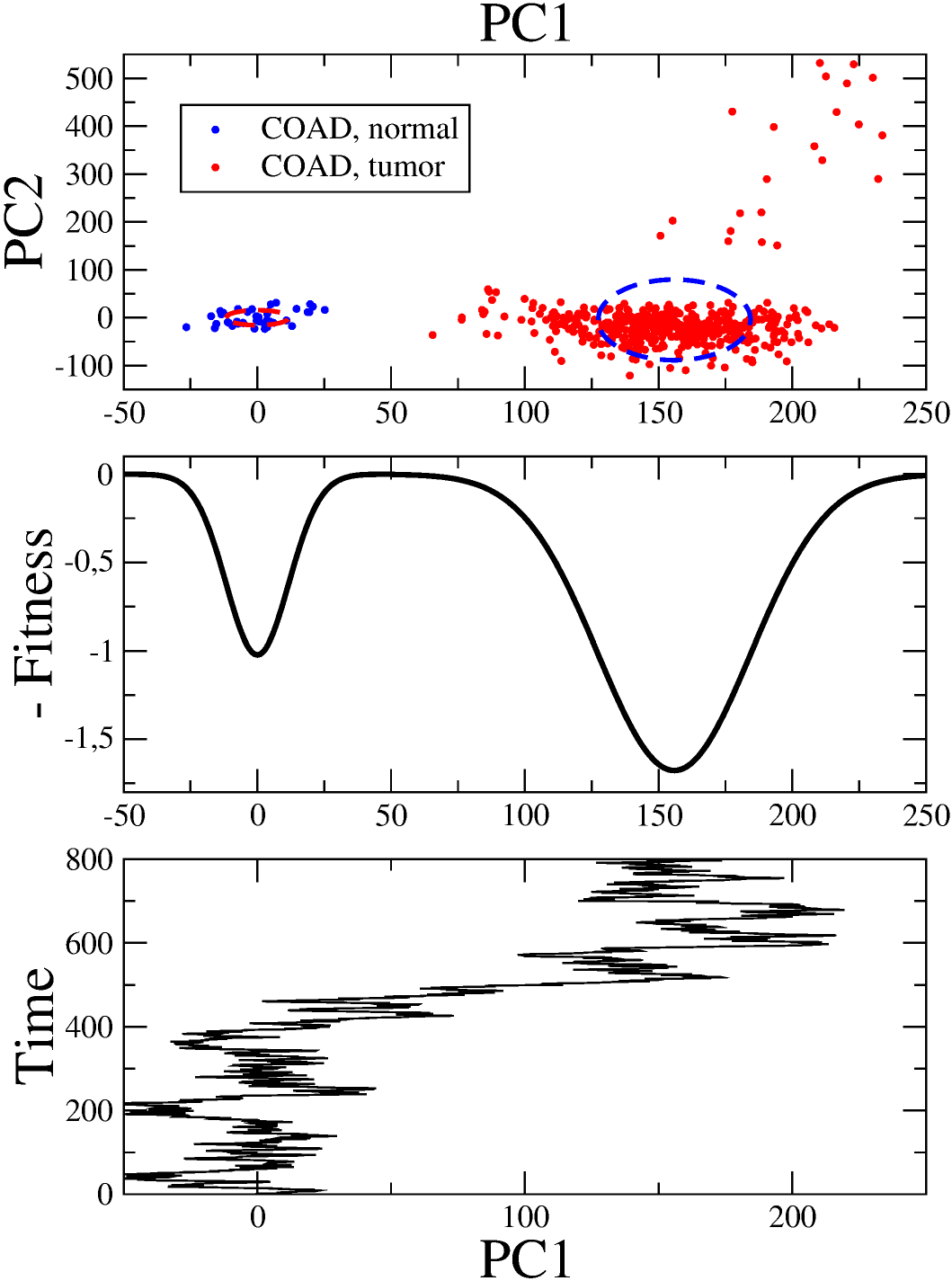}
\caption{Top: A PCA representation of the GE data in COAD. Ellipses illustrate the magnitudes
of the r.m.s. radii along each direction. Center: Schematic representation of fitness along
the PC1 axis. Bottom: Time evolution of a microstate which starts in the normal region and
transits to the tumor region.}
\label{fig1}
\end{center}
\end{figure}

A useful picture of normal and tumor tissues comes from processing gene expression (GE) data for
small tisue samples \cite{Landscape}. A small sample is conformed by the contribution of many 
interacting cells, and is represented by a point in GE space, a tissue microstate. Recall,
for example, the GE data provided by The Cancer Genome Atlas \cite{TCGA} for colon adenocarcinoma
(COAD). A principal component (PC) \cite{PCA} representation is shown in the top panel of Fig. \ref{fig1}.
Two clouds of points are clearly distinguished. The normal samples are distributed around the
origin of coordinates. The dispersion of points is related to regional differences and the fact
that the samples come from different patients. An ellipse representing the r.m.s. radii along
each axis is drawn. The tumor samples, on the other hand, are far apart along the PC1 axis.
An ellipse with the r.m.s. cloud radii is also drawn. Let us denote by $\bar x_1$, $R_n$, and
$R_t$ the position along PC1 of the center of the tumor cloud, and the r.m.s. radii along PC1
of the normal and tumor clouds, respectively. It is apparent that the distance between cloud 
centers, $\bar x_1=155.89$, is much larger than any of the radii $R_n=11.71$ or $R_t=28.53$. 
 
Both the normal and tumor regions may be qualified as attractors in GE space \cite{attractors}. 
they may sustain long time oscillations of microstates. Their radii $R_n$ and $R_t$ define
their basins of attraction.

Normal and tumor attractors could be understood as global stable solutions of GE networks
\cite{networks}. We prefer, however, to interpret them as local maxima in the fitness landscape.
The center panel of Fig. \ref{fig1} contains a sketch of the fitness along the PC1 axis. The
$y$ axis is the fitness with a minus sign. The centers of the clouds correspond to local 
maxima. We computed the volumes of the basins of attraction, which provide an estimation
of the number of available microstates \cite{entropy}. This number is much higher for tumors
than for normal samples. In addition, the replication rates of tumor stem cells is usually
larger than the rate of normal cells \cite{rates}. The conclusion is that the tumor minimum
is the deepest in Fig. \ref{fig1} central panel, i.e. the one with the highest fitness. 
The intermediate region, $R_n<x_1<\bar x_1-R_t$, correspond to a low fitness barrier
separating the wells. The scarcity of samples in this region is an evidence of the low
fitness values. The barrier is needed in order to prevent the spontaneous transitions from
the normal to the tumor regions, driven by the fitness difference. 

The PC1 axis describes progression to cancer. A normal sample evolving to a tumor state
realizes a motion along the PC1 axis from a point near the origin to a point near $\bar x_1$.
We have schematically drawn its motion in the bottom panel of Fig. \ref{fig1}. 

The random displacements, i.e. random variations in the GE of a gene or a group of genes,
are caused by spontaneous somatic mutations \cite{somatic}, epigenetic changes \cite{epigenetic}
or the action of external factors \cite{carcinogens}. Because the normal region is a local
maximum of fitness, the sample may experience random displacements, but remain in the normal region
for a long time. Only when the random displacements drive the sample out of the normal
region it may experience a drift towards the tumor zone. Notice that the distance between
normal and tumor regions is $R=\bar x_1-R_n-R_t$. This is the minimal length of a path
connecting both regions.

Among the random motions in GE space, we distinguish small (Brownian) displacements and large 
(Levy) jumps. The former are most likely caused by independent gene variations, whereas the
latter represent the coordinated variations of a set of genes. Levy jumps are described 
by power-like distribution functions, which are very common in mutations \cite{LevyMut}, and
in the GE distribution functions of eukaryotes \cite{Eukaryotes,GErearr}. The Levy jumps could
perhaps be identified as the hits in the multistep theory of cancer \cite{multistep}.

In the bottom panel of Fig. \ref{fig1}, a trajectory which combines Brownian motions and large jumps
is drawn. However, the parameters (fluctuation scale and the rate of large jumps) have been 
exagerated in order to make the effects apparent.

Under the assumption that Levy jumps are rare (unfrequent) events, we compute the one-hit
probability to transit from the normal to the tumor region, leading to the following 
expression for the risk \cite{LevyCancer}:

\begin{equation}
\frac{risk}{N_{s}} = \mu \frac{D}{R} (t_0+m_{s}\times age).
\label{risk}
\end{equation}

\noindent
The parameter $D$ sets the scale for fluctuations. $\mu$ is the effective rate of large jumps. 
In an effective way, it accounts for the effects of hereditary factors, external carcinogens, 
the role of the immune system, etc. $t_0=\log_2 N_{s}$. For the set of tissues considered in Refs.
\cite{Tomasetti1,Tomasetti2}, $t_0$ is a number around 30. On the other hand,  
$m_{s}\times age$ is the number of stem cell generations measured from the time $t_0$ at which the
tissue is formed. For many tissues, the product $m_s\times age$ is larger than 30, thus appart from 
nearly constant magnitudes, $risk\sim N_s\times m_s$.

The analysis in the next section is based on Eq. (\ref{risk}), or its simpler version $risk\sim N_s\times m_s$. 

\section*{Results}

%Up to three levels of \textbf{subheading} are permitted. Subheadings should not be numbered.

\subsection*{Groups of tissues and risk of cancer}

Table \ref{tab1} contains a summary of the compiled data \cite{Tomasetti1,Tomasetti2}. 
The data shows relatively large variations. For example, the stem-cell division rate in
colon crypts is reported by the authors to be 73/year, whereas for bronchio-alveolar cells
the rate is around 0.07/year, i.e. three orders of magnitude smaller. Similar variations 
are exhibited by the number of stem cells or the quotient $N_s/N_{cell}$ between the stem 
and total cell numbers in the tissues. 

\begin{figure}[t]
\begin{center}
\includegraphics[width=0.9\linewidth,angle=0]{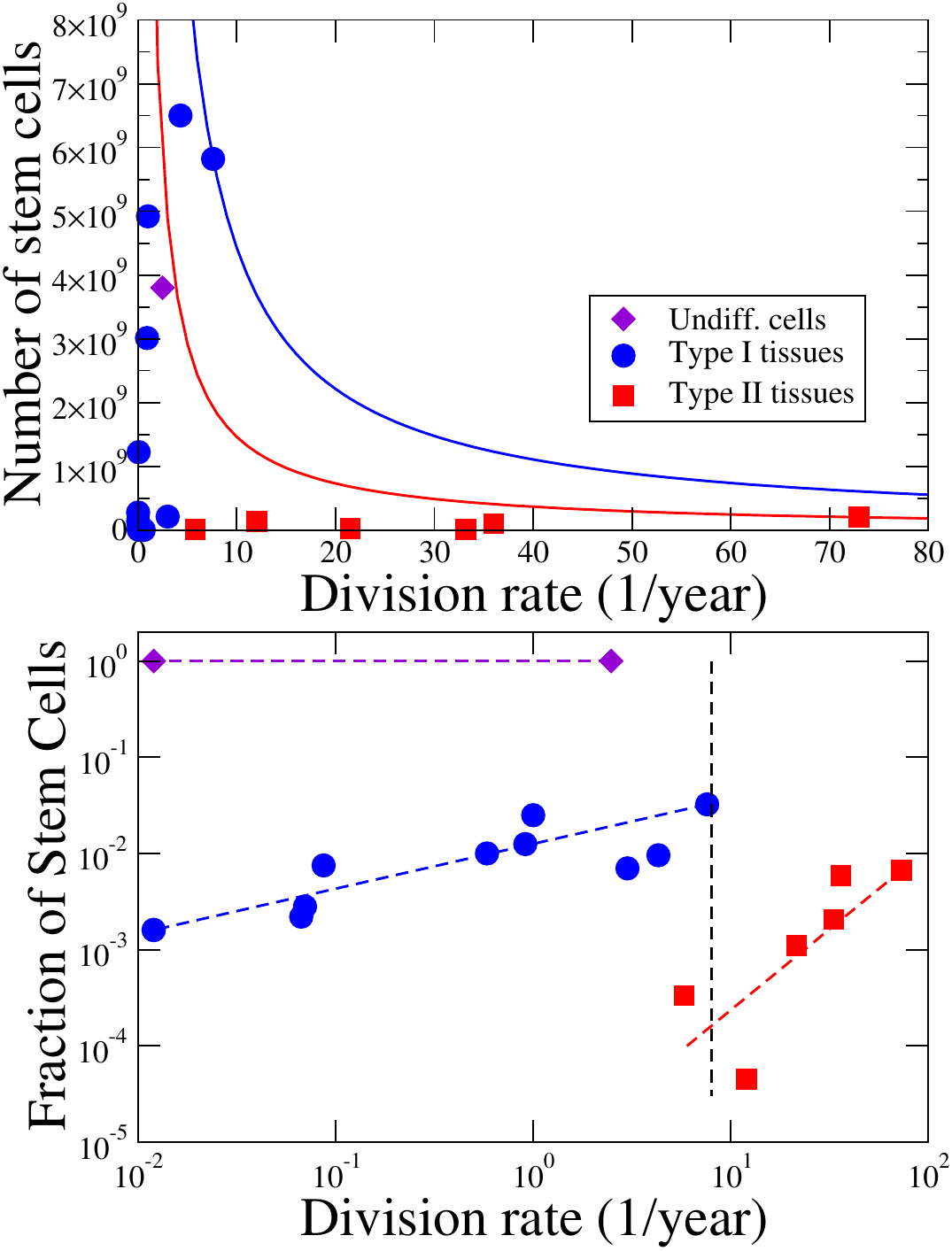}
\caption{Top: Representation of tissues from Table \ref{tab1} as points in the $(m_s,N_s)$ plane. 
Hyperbolae containing the colon (red) and the epidermis (blue) are drawn as illustrations. 
Bottom: A log-log plot of $N_s/N_{cell}$ vs $m_s$, where tissue clustering becomes more apparent. 
Dashed lines are guides to the eyes. The vertical one corresponds to $m_s=8$/year.}
\label{fig2}
\end{center}
\end{figure}

In Fig. \ref{fig2}, upper panel, we represent tissues as points in the $(m_{s},N_{s})$ plane. 
Because of Eq. (\ref{risk}), tissues with similar risks of cancer should be located roughly along 
a hyperbola with a given value of the product $N_{s}\times m_{s}$. If there is a maximal allowed
risk, then the corresponding hyperbola would divide the quadrant in allowed and forbidden regions. 
An increase of the division rate should be compensated by a decrease in the stem cell number, 
and viceversa, in order to keep the risk below the maximum. Thus, we can distinguish tissues 
with relatively ``high division rates'' (and low stem cell numbers) or ``high stem cell numbers'' 
(and low division rates).
 
In Fig. \ref{fig2}, lower panel, logarithmic scales are used in order to make apparent the groups of 
tissues. First, we have the undifferentiated cells (melanocytes, ovarian germ cells), represented by 
diamonds, for which $N_{s}/N_{cell}=1$. 
Next, there is a group we call Type I tissues (circles), characterized by a relatively high fraction 
of stem cells and division rates lower than around $8$/year. They correspond mainly to interior organs, 
although the epidermis also belongs to this group. Finally, we have Type II tissues (squares) with 
relatively high division rates, $m_s\ge 8$/year, and low fractions of stem cells.

This lower panel shows that each group (I and II) exhibits a monotonous dependence of $N_{s}$ on $m_{s}$. 
In the next subsection, we argue that this is indeed a parametric dependence. Different tissues inside 
a group differ in the average replacement rate of their differentiated cells. As this parameter is increased, 
both $N_{s}$ and $m_{s}$ increase, leading to the observed monotony.

The enigmatic threshold at $m_{s}\approx 8$/year separating the two groups, where the fraction of stem 
cells falls down two orders of magnitude, could be understood in terms of a maximal allowed risk of 
cancer dictated by evolution. A tissue in Group I could not go beyond the extremal point (the skin, 
which defines the extremal hyperbola), thus a tissue with a higher average replication rate could be 
realized only as a Group II tissue.

Notice, for example, that the risk of cancer in the male germinal cells, first point in the lower curve, 
is reduced by two orders of magnitude as compared to the risk in the skin, which has a similar value of 
$m_{s}$. 

The maximal allowed value for the risk could be estimated as half the observed risk for basal cell carcinoma.
Indeed, the reported value of 0.3 \cite{Tomasetti1}, corresponds to an age equal to 80 years.
The biological age for our species is, however, only around 40 years \cite{age}, 
thus according to Eq. (\ref{risk}), the maximal allowed risk in human tissues should be around 0.15.   

\subsection*{Tissue groups and homeostasis}

In this subsection, we consider the tissue under stationary conditions, i.e. homeostasis, and 
examine the equations determining the number of cells. We shall see that these equations offer 
another point of view to the origin of the group of tissues mentioned above.

\begin{figure}[ht]
\begin{center}
\includegraphics[width=0.9\linewidth,angle=0]{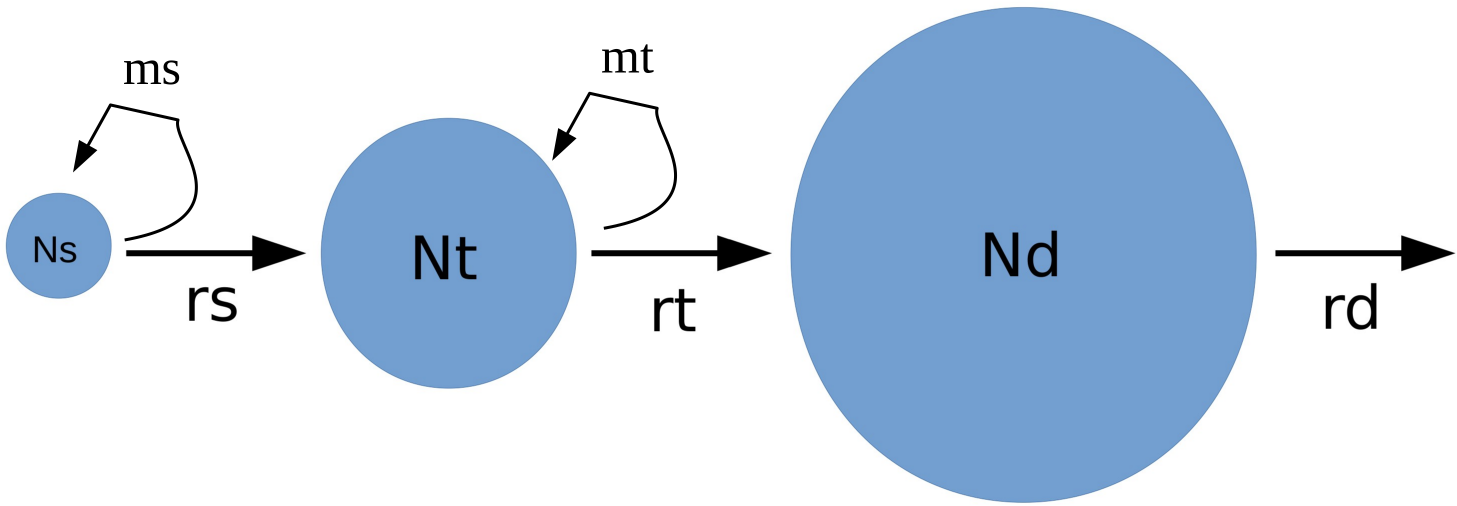}
\caption{Schematic representation of homeostasis in a tissue. See explanation in the main text.}
\label{fig3}
\end{center}
\end{figure}

We assume that, in addition to stem, in the tissue there are a number $N_t$ of transit amplifying cells, 
and a number $N_d$ of fully differentiated cells. We schematically represent in Fig. \ref{fig3} the 
equilibrium equations. The diameters of circles in the figure indicate that $N_d>>N_t>>N_s$. Arrows 
represent rates of division ($m_s$, $m_t$), differentiation ($r_s$, $r_t$) or replacement ($r_d$).

Under homeostatic conditions, the number of cells are nearly constant, meaning that fluxes should be equilibrated:

\begin{eqnarray}
m_{s} N_s&=&r_{s} N_s,\nonumber\\
r_{s} N_{s}+m_{t} N_{t}&=&r_{t} N_{t},\\
r_{t}N_{t}&=&r_{d}N_{d}.\nonumber
\label{eq1}
\end{eqnarray}

\noindent
The first equation leads to $m_{s} =r_{s}$. The next two may be rewritten as:

\begin{eqnarray}
r_{t} -m_{t} &=& \frac{N_s}{N_{t}} m_s << m_s,\\
r_{t}&=& \frac{N_{d}}{N_t} r_d >> r_d.\nonumber
\label{eq2}
\end{eqnarray}

\noindent
In addition, it is expected that the division rate of transit cells is much higher than that of stem cells, 
thus $r_t$ and $m_t$ are very close and we arrive to the conditions: $m_t\approx r_t>> m_s$, and $m_t>>r_d$.

The above equations may help understanding the groups apparent in Fig. \ref{fig2} lower panel. Inside a group, 
we expect an increase of $m_s$ when $r_d$ is increased. This does not follow from the equations, but is a very 
plausible hypothesis. With regard to the fraction of stem cells, we may write an equation for $N_s/N_d$: 

\begin{equation}
\frac{N_s}{N_d}=\frac{r_d}{m_s} (1-m_t/r_t).
\label{eq3}
\end{equation}

\noindent
As mentioned. the left hand side of the equation should be less than one. The difference between the two 
groups comes from the expression in parenthesis. In the second group of tissues this expression takes a value 
much closer to zero, because $m_t$ and $r_t$ simultaneously take very large values, and their ratio is 
very close to one.

We may examine some of the studied tissues in quality of examples. Let us start with the epidermis, the one with 
the highest division rate in Type I tissues, with $m_s=7.6/$year. Authors of Ref. \cite{Tomasetti1} provide the 
following estimations: $N_s/N_{cell}\approx 0.032$, $N_t/N_{cell}\approx 0.29$, from which it follows that 
$(N_t+N_s)/N_{cell}\approx 0.32$. This high proportion of maintenance cells seems to be the distinctive characteristic 
of Type I tissues. From these numbers and the replacement rate $r_d\approx 18/$year \cite{rd1,rd2,rd3}, we get  
$r_t\approx 42$/year and $m_t\approx 41/$year. 

On the other hand, let us consider the Type II tissue with the highest division rate: the colon. Because of the 
derived relation,  $m_t >> m_s\approx 73/yr$, we expect $m_t$ to be around 200/year or even higher. Notice that 
there is a characteristic value for the division rate of fast replicating human tissues \cite{limit}, of around 
one division per day or $400/yr$. 

In order to verify that this characteristic, a high $m_t$ close a circadian variation, is distinctive of Type II 
tissues, we examine blood, which is in the opposite ``low-$m_s$'' side. Blood , with $m_s\approx 12$/year, exhibits 
a complex pattern of differentiation channels. The replacement rate of neutrophils, for example, is 
$r_d\approx 73 - 365/$year \cite{rd1,rd4}. Because of the condition $m_t >> r_d$, we get again a very high value for 
$m_t$ in the neutrophil line.

\section*{Concluding remarks}

According to data provided by papers \cite{Tomasetti1,Tomasetti2}, and their suggestion that the product 
$N_s\times m_s$ (multiplied by the age) is an indicator of the risk of cancer in the tissue, 
further supported by Eq. (\ref{risk}), we suggest a classification of tissues dictated by their position 
in the ($m_s$,$N_s$) plane.

Type I, i.e. ``high-$N_s$, low-$m_s$'', tissues are characterized by a high proportion of maintenance cells, 
reaching one third. Type II, i.e. ``high-$m_s$, low-$N_s$'' tissues, on the other hand, exhibit very high 
values for the division rates of transit amplifying cells, reaching 400 divisions per year.

Each group of tissues conform a well defined cluster, as shown in Fig. \ref{fig2} lower panel. The equations 
determining the numbers of cells in a tissue under homeostatic conditions allow a qualitative understanding of 
the two groups in terms of the fraction of stem to total number of cells and the division rate of transit 
amplifying cells. The threshold value $m_s\approx 8/$year, separating the two groups, could be dictated by 
an upper bound impossed on the risk by evolution, and the possibility of a second solution of the master 
equations determinig the number of cells.

%\section*{Methods}

%Topical subheadings are allowed. Authors must ensure that their Methods section includes adequate experimental and characterization data necessary for others in the field to reproduce their work.

%\bibliography{sample}

%\noindent LaTeX formats citations and references automatically using the bibliography records in your .bib file, which you can edit via the project menu. Use the cite command for an inline citation, e.g.  \cite{Hao:gidmaps:2014}.

%For data citations of datasets uploaded to e.g. \emph{figshare}, please use the \verb|howpublished| option in the bib entry to specify the platform and the link, as in the \verb|Hao:gidmaps:2014| example in the sample bibliography file.

\section*{Acknowledgements}

A.G. acknowledges the Cuban Program for Basic Sciences, the Office of External Activities of 
the Abdus Salam Centre for Theoretical Physics, and the University of Electronic Science and 
Technology of China for support. The research is carried on under a  project of the Platform 
for Bioinformatics of BioCubaFarma, Cuba. 

\section*{Author contributions statement}

A.G. is responsible for the analysis based on master equations and redaction of an initial draft.  
All authors analyzed the data and reviewed the final version of the manuscript. 

\section*{Additional information}

%To include, in this order: \textbf{Accession codes} (where applicable); 

\textbf{Competing interests} The authors declare that there are not competing interests. 

%The corresponding author is responsible for submitting a \href{http://www.nature.com/srep/policies/index.html#competing}{competing interests statement} on behalf of all authors of the paper. This statement must be included in the submitted article file.

\begin{table}[b]
\centering
\begin{tabular}{|c|c|c|c|c|}
\hline
 Undifferentiated cells & $N_{cell}$ & $N_s$ & $m_s$ (1/yr)\\
 \hline
 {Melanocytes} & 3.8 x $10^9$ & 3.8 x $10^9$ & 2.48\\
 {Ovarian germinal cells} & 1.1 x $10^7$ & 1.1 x $10^7$ & $< 0.012$\\
\hline
 Type I tissues & $N_{cell}$ & $N_s$ & $m_s$ (1/yr)\\
 \hline
 {Epidermis}  & 1.8 x $10^{11}$ & 5.82 x $10^9$ & 7.6 \\
 {Breast}  & 6.8 x $10^{11}$ & 6.5 x $10^9$ & 4.3 \\
 {Prostate}  & 3 x $10^{10}$ & 2.1 x $10^8$ & 3 \\
 {Pancreas}  & 1.97 x $10^{11}$ & 4.92 x $10^9$ & 1 \\ 
 {Hepatocytes}  & 2.41 x $10^{11}$ & 3.01 x $10^9$ & 0.9125 \\
 {Gallbladder}  & 1.6 x $10^8$ & 1.6 x $10^6$ & 0.584 \\
 {Thyroid}  & 1.1 x $10^{10}$ & 8.25 x $10^7$ & 0.087 \\
 {Bronchio alveolar cells}  & 4.34 x $10^{11}$ & 1.22 x $10^9$ & 0.07 \\
 {Bone cells}  & 1.9 x $10^9$ & 4.18 x $10^6$ & 0.067 \\
 {Cerebellum}  & 8.5 x $10^{10}$ & 1.36 x $10^8$ & $< 0.012$\\
 {Brain}  & 1.707 x $10^{11}$ & 2.73 x $10^8$ & $< 0.012$\\
 \hline
 Type II tissues & $N_{cell}$ & $N_s$ & $m_s$ (1/yr)\\
 \hline
 {Colon}   & 3 x $10^{10}$ & 2 x $10^8$ & 73 \\
 {Small Intestine}   & 1.7 x $10^{10}$ & 1 x $10^8$ & 36 \\
 {Esophagus}   & 3.24 x $10^9$ & 6.65 x $10^6$ & 33.18 \\
 {Head and Neck}   & 1.67 x $10^{10}$ & 1.85 x $10^7$ & 21.5 \\
 {Blood cells}   & 3 x $10^{12}$ & 1.35 x $10^8$ & 12\\
 {Germ cells Testis}   & 2.16 x $10^{10}$ & 7.2 x $10^6$ & 5.8 \\
 \hline
\end{tabular}
\caption{Data, coming from Refs. \cite{Tomasetti1,Tomasetti2},
for different human tissues.}
\label{tab1}
\end{table}

\end{document}